\documentclass[twocolumn]{IEEEtran}
\pdfoutput=1

\usepackage[utf8]{inputenc}
\usepackage{amsmath}
\usepackage{amsthm}
\usepackage{amsfonts}
\usepackage{pifont}
\usepackage{amssymb}
\usepackage{color}
\usepackage{bm}
\usepackage[hyphens]{url}
\usepackage{graphicx}
\usepackage{cite}
\usepackage{float}

\usepackage[caption=false,font=footnotesize]{subfig}

\usepackage{balance}

\definecolor{OliveGreen}{rgb}{0,0.6,0}

\newcommand{\paperTitle}{Monitoring of Underwater Critical Infrastructures: the Nord Stream and Other Recent Case Studies\vspace{1.2mm}}

\allowdisplaybreaks

\begin{document}

\title{\paperTitle}

\author{Giovanni Soldi, Domenico Gaglione, Simone Raponi, Nicola Forti, Enrica d'Afflisio,  
Pawe\l{} Kowalski, \\  Leonardo M. Millefiori, Dimitris Zissis, Paolo Braca, Peter Willett, Alain Maguer, Sandro Carniel, \\ Giovanni Sembenini, and Catherine Warner

\thanks{G. Soldi, D. Gaglione, S. Raponi, N. Forti, E. d'Afflisio, P. Kowalski, L. M. Millefiori, P. Braca, A. Maguer, S. Carniel, G. Sembenini, and C. Warner are with the NATO STO Centre for Maritime Research and Experimentation (CMRE), La~Spezia, Italy (e-mail: \{name.surname\}@cmre.nato.int)}  %
\thanks{D. Zissis is with the University of the Aegean, 84100 Syros, Greece (e-mail: dzissis@aegean.gr).}
\thanks{P.\ Willett is with the University of Connecticut, Storrs, CT 06269, USA (e-mail: peter.willett@uconn.edu).}
\thanks{Financial support from the Office of the NATO Chief Scientist and the NATO Allied Command Transformation (ACT) is greatly acknowledged.}
\thanks{The views expressed in this research paper are the responsibility of the authors and do not necessarily
reflect the opinions of the NATO.}
}
		
\maketitle

\begin{abstract}
The explosions on September 26th, 2022, which damaged the gas pipelines of Nord Stream~1 and Nord Stream~2, have highlighted the need and urgency of improving the resilience of Underwater Critical Infrastructures (UCIs).
Comprising gas pipelines and power and communication cables, these connect countries worldwide and are critical for the global economy and stability.
An attack targeting multiple of such infrastructures simultaneously could potentially cause significant damage and greatly affect various aspects of daily life.
Due to the increasing number and continuous deployment of UCIs, existing underwater surveillance solutions, such as Autonomous Underwater Vehicles (AUVs) or Remotely Operated Vehicles (ROVs), %
are not adequate enough to ensure thorough monitoring.

We show that the combination of information from 
both underwater and above-water surveillance sensors %
enables achieving Seabed-to-Space Situational Awareness (S3A), mainly thanks to %
Artificial Intelligence (AI) and Information Fusion (IF) methodologies. These are designed to process immense volumes of information, fused from a variety of sources and generated from monitoring a very large number of assets on a daily basis. %
The learned knowledge can be used to anticipate future behaviors, identify threats, and determine critical situations concerning UCIs.  

To illustrate the capabilities and importance of S3A, we consider three events that occurred in the second half of 2022: the aforementioned Nord Stream explosions,
the cutoff of the underwater communication cable SHEFA-2 connecting the Shetland Islands and the UK mainland, and the suspicious activity of a
large vessel in the Adriatic Sea. Specifically, we provide
analyses of the available data, from Automatic Identification System (AIS) and satellite data, integrated with possible contextual information, e.g., bathymetry, Patterns Of Life (POLs), weather conditions, and human intelligence (HUMINT). 
 
\end{abstract}

\section{Introduction}
\label{sec:introduction}
\begin{figure}
    \centering
    \includegraphics[width=.99\columnwidth]{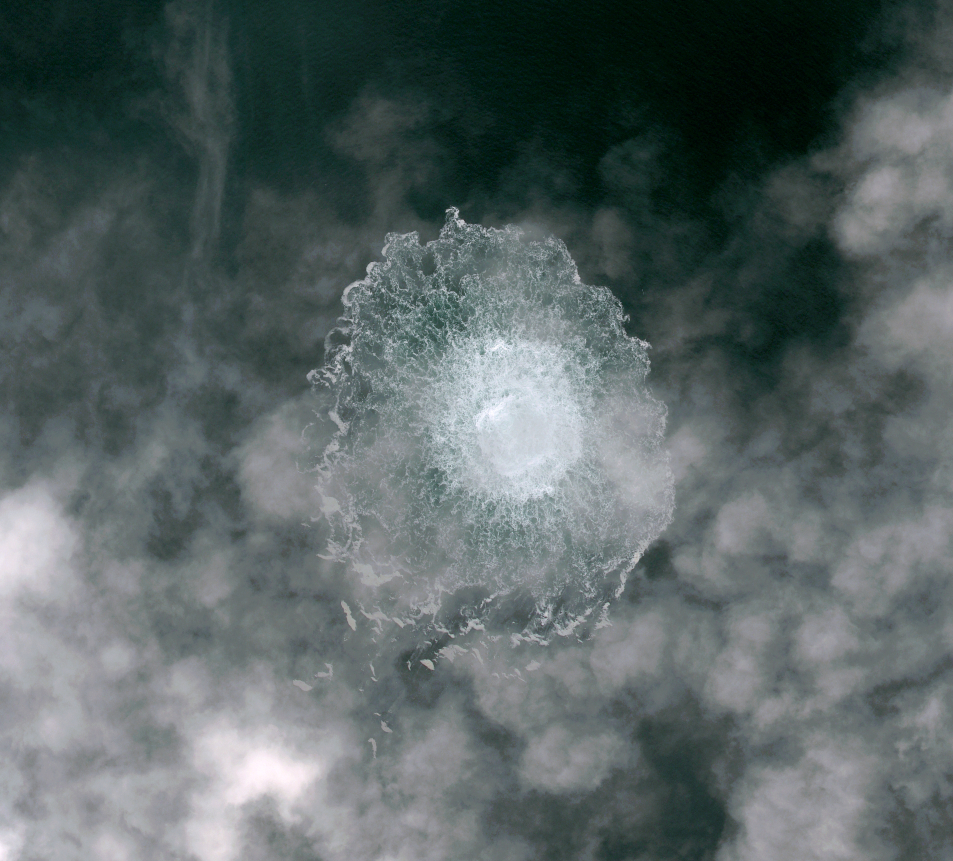}
    \vspace{-2mm}
    \caption{Nord Stream leak as captured by Pl\'eiades Neo (ESA \copyright~Pl\'eiades Neo).
    The diameter of the leak is estimated in 0.5--0.7 km \cite{JIA2022100210}.}
    \label{fig:my_label}
\end{figure}

On September 26th, 2022, Danish and Swedish seismometers detected~\cite{nord-stream-press} a series of explosions on the Nord Stream~1 and Nord Stream~2 underwater gas pipelines. These explosions, besides causing severe damage to the pipes, 
led to three underwater gas leaks, with the subsequent release  of %
an enormous amount of methane into the atmosphere. Figure~\ref{fig:my_label} shows an image of the methane leak acquired by the satellite constellation Pl\'eiades Neo, made available by ESA. %

\begin{figure*}[!t]
    \centering
    \includegraphics[width=.8\textwidth]{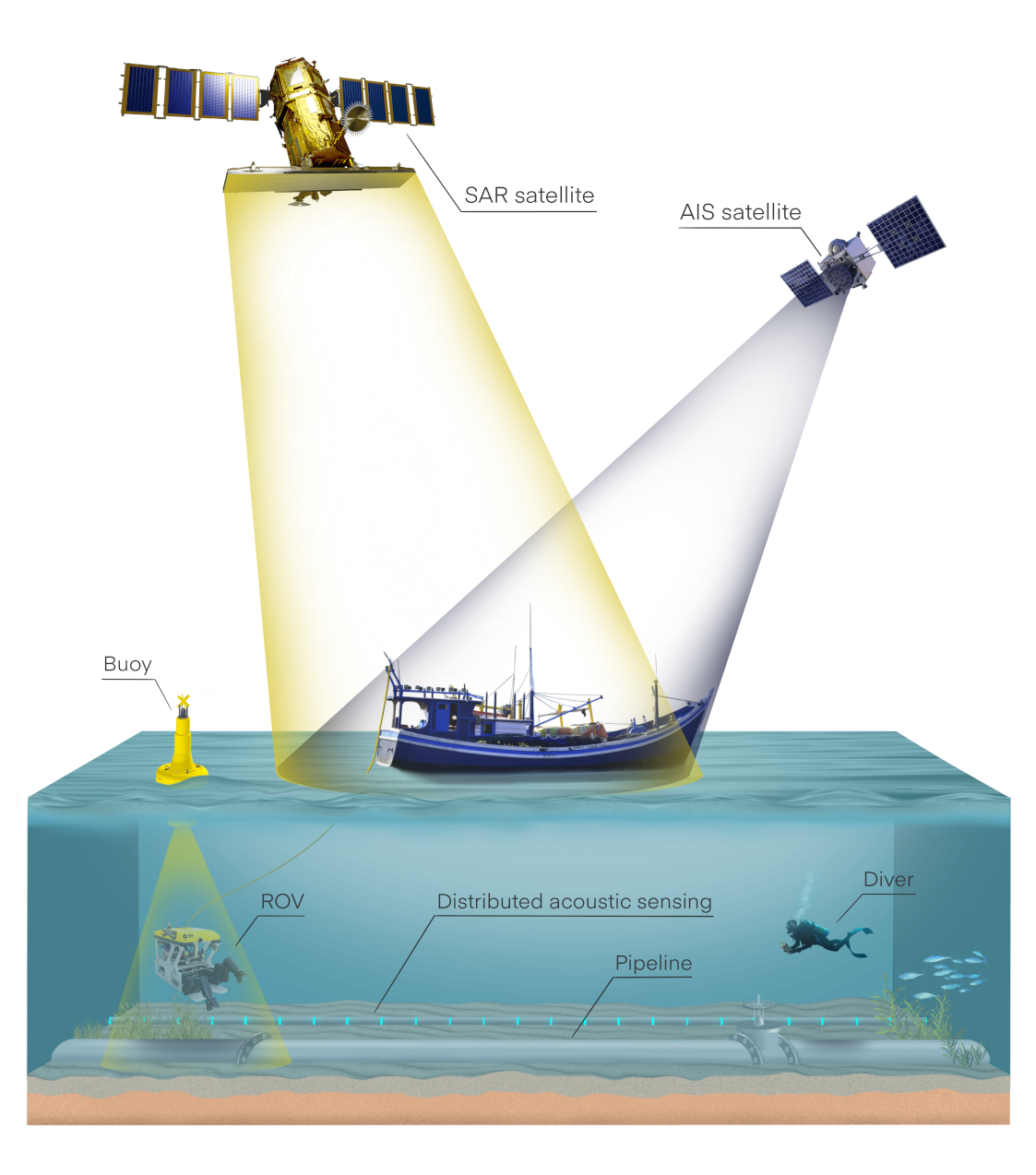}
    \vspace{-2mm}
    \caption{Conceptual illustration of the monitoring of an UCI. Depending on the location of the UCI and weather conditions, an above-water sabotage might be potentially conducted by a specialized diver supported by a surface vessel; a remotely operated vehicle (ROV)
    may also be employed in some circumstances.
    Above-water systems and sensors, such as the Automatic Identification System (AIS) and Synthetic Aperture Radar (SAR), may provide large-scale monitoring capabilities. 
    These may be complemented by underwater assets, such as Distributed Acoustic Sensing (DAS), to ensure a comprehensive maritime surveillance.}
    \label{fig:maritime-scenario}
\end{figure*}

Despite suspicions of sabotage from certain authorities and organizations, particularly given the current political climate in Eastern Europe, there is currently no concrete evidence to indicate how and by whom the explosions were caused.
At the same time,
these events have also put the whole climate community on alert, due to the  much more powerful greenhouse effect (approximately 30 times) of methane compared to carbon dioxide, especially in the short term. It has been estimated that more than 220,000 tonnes of methane, comparable
to the
annual anthropogenic methane emissions in Austria, had probably been released in the atmosphere during the Nord Stream leakage~\cite{JIA2022100210}.  
While the accident per se is not significantly changing the figures of greenhouse gas emissions leading to global warming and climate change, it nevertheless represents an unprecedented case of interlink between climate change and security aspects, that needs to be properly accounted for by governments.

The Nord Stream incident has brought attention to the vulnerability of Underwater Critical Infrastructures (UCIs) such as gas pipelines and underwater cables. This has led to increased focus from both the public and policymakers on improving the resilience of these vital assets, as there is growing concern that similar malicious operations will occur in the future.
Indeed, recently, on December 22nd, 2022, Italian newspapers reported~\cite{adriatic-sea-press} the suspicious activity of a large vessel in close proximity to the Trans Adriatic Pipeline (TAP) in the Adriatic Sea.
Therefore, the protection and surveillance of UCIs are crucial elements that will be included in any future maritime strategy.

Sabotage of UCIs can occur through the use of surface assets such as warships or commercial vessels, or by underwater assets, as shown in Fig.~\ref{fig:maritime-scenario}.
In the former case,
multiple above-water heterogeneous systems and 
sensors, e.g., Automatic Identification System (AIS), satellite sensors~\cite{Space_1_AESM}, 
and terrestrial radars, may have a crucial role to provide a seamless %
large-scale
Maritime Surveillance (MS), even in remote areas of the world.
In the latter case, underwater sensors (such as active/passive sonars and cameras) installed on the UCIs~\cite{cables_monitoring_1}, or equipped on Unmanned Underwater Vehicles (UUVs)~\cite{cables_monitoring_2}, would complement MS by providing undersea monitoring capabilities. Besides sensory data, the analysis of contextual information, such as bathymetry, weather data, human intelligence (HUMINT), and open source intelligence (OSINT), is of paramount importance. More in general, the joint use of underwater and above-water heterogeneous sensors, together with contextual and intelligence information is a key concept for the transition towards a Seabed-to-Space Situational Awareness (S3A). 
Given the scale of the problem and %
the large amount of data to be processed, achieving S3A related to the monitoring of UCIs can be done by using advanced Artificial Intelligence (AI) and Information Fusion (IF) techniques. %
These techniques allow for the integration of vast amounts of information from various sources and monitoring a large number of assets on a daily basis.
Examples of such techniques include Bayesian Multi-Target Tracking (MTT) techniques~\cite{Space_2_AESM, Fusion_AIS, GagSolMeyHlaBraFarWin:J20}, multi-reasoning systems based on the Dempster-Shafer theory~\cite{KOWALSKI202329}, and anomaly detection techniques. These enable the fusion of diverse information at multiple levels. %
The knowledge gained from these techniques, once extracted and made easily understandable, can provide end-users such as government authorities, defense forces, coast guards, and police, not only with a better understanding of the entities and actors involved in a specific event, such as the Nord Stream incident, their relations, and the potential consequences of these relations, but also with an effective tool to anticipate future threats to UCIs and other critical assets.
We anticipate that these techniques will provide the ability to prevent potential future attacks on UCIs and will become of increasing interest to national and international institutions and agencies, as well as the maritime industry.

The remainder of this paper is organized as follows: Section~\ref{sec:uci} provides an overview of UCIs and describes important aspects related to their resilience. Section~\ref{sec:sensors_contextutal_information} provides a description of sensor technologies
that could effectively be used for monitoring UCIs, and details useful contextual information.
Section~\ref{sec:seabed-to-space-situational-awareness} provides an overview of S3A and, in particular, it describes state-of-the-art information fusion, anomaly detection, and automated reasoning techniques. Section~\ref{sec:use_cases} presents
analyses
on the Nord Stream explosions,
the cutoff of the underwater cable SHEFA-2 connecting the Shetland Islands and the UK mainland~\cite{shetland-accident-press-1}, and the anomalous behaviour of a
large vessel in the Adriatic Sea.
Concluding remarks are provided in Section~\ref{sec:conclusion}.

\section{Underwater Critical Infrastructure}
\label{sec:uci}

Nowadays, submarine pipelines and cables represent a vital infrastructure %
for global finance, economy, maritime security, and everyday life. 
Due to their undersea concealment, concern about UCIs is usually risen to the public and institutional attention only after a major accident occurs, a phenomenon referred to in~\cite{Bueger22} as ``collective sea blindness''.
Despite their importance and influence on many aspects of %
our societies, UCIs have only recently seen increasing political and scholarly attention, mostly sparked by military concerns and recent accidents involving underwater gas pipes and cables. 
Submarine pipelines %
are the backbone for energy transportation to the market, e.g., 
oil and gas, as they connect increasingly complex structures, such as offshore rigs, 
floating storage, and floating processing units, that directly feed ashore. 
The pipes themselves are of steel, and concrete is used to prevent impacts and damage from ships’ anchors. But these resources are nonetheless vulnerable and thence of concern, witness the recent explosion of the Nord Stream pipeline and its effect on public concern about the resilience of these important infrastructures. Similarly, the submarine cable network, with more than 400 active cables spanning at least 1.3 million kilometers globally, is a vital asset composed of optical fibers and energy cables laid on the ocean floor. They constitute the most efficient and cost-effective solution to sending digital information across the globe.
This network digitally connects countries worldwide, with more than 10 trillion USD dollars in financial transactions exchanged every day, and represents the backbone for internet communications globally. 

Compared to underwater energy pipelines, underwater cables are more vulnerable, since they are more flexible and fragile.
The major causes of damage are represented by human errors and negligence. To make an example, 40\% of the incidents occur due to trawling activities by fishing boats, while another 15\% is caused by anchoring accidents, such as improperly stored anchors, anchoring outside approved areas, anchoring mispositioning due to weather conditions, and emergency dropping of an anchor.
Other human-driven benign causes of incidents include oil and gas development, offshore wind and energy constructions, hydro-energy projects, and deep-sea mining operations.
Intentional sabotage operations to underwater cables classified as hybrid warfare operations have not yet been officially documented. However, different Russian submarine activities in the proximity of underwater cables have been publicly reported since 2015, raising concern among NATO officials, as the Russian navy has clearly demonstrated an unprecedented interest in undersea cables~\cite{Euractiv_article}.
In February 2022, Russia has conducted a naval exercise just outside Ireland's exclusive economic zone, very close to several submarine cables connecting France, the UK, and the USA. According to an Irish military source, the scope of this operation was a demonstration of capability to sabotage underwater cables~\cite{Irish_times}.

The highest risk associated with underwater cables is represented by high-density maritime bottlenecks. For instance, seven intercontinental cables pass through the Strait of Gibraltar, between Morocco and the Iberian Peninsula, connecting the Mediterranean Sea with the Atlantic Ocean. Another critical point is the passage through the Red Sea between the Mediterranean Sea and the Indian Ocean, with sixteen underwater cables passing through the Egyptian mainland~\cite{Bueger22}.
Even though most countries could cope with a significant decrease in bandwidth in case of simultaneous damage to multiple underwater cables, island states and oversea territories, not boasting the same redundancy, are the most vulnerable ones.
A recent example is the interruption of communications 
with the Shetland Islands and the Faroe Islands, %
after the communication cable SHEFA-2 was severely damaged in two distinct points, most likely by trawling fishing boats. 

The aforementioned critical aspects of the underwater cable networks are analyzed in terms of graph robustness and resilience in the next subsection.  

\subsection{Robustness and resilience}
\label{subsec:robustness_resilience}

UCIs are increasingly interconnected and interdependent, thus providing valuable benefits in terms of efficiency, quality of service, performance, and cost reduction. However, these interdependencies increase the vulnerability of UCIs to accidental and malicious threats, as well as the risk of a domino effect on the whole networked infrastructure. Consequently, the impact of infrastructure components' failures can be aggravated and more difficult to predict, compared to failures confined to a single infrastructure. As an example, blackouts can be caused by the outage of a single transmission element not properly managed by automatic control actions or operator intervention, gradually leading to cascading outages and eventually to the collapse of the entire network. 
Examples of cascading effects due to infrastructure interdependencies leading to catastrophic events across multiple infrastructures spanning wide geographical areas are documented in~\cite{pourbeik06}. 

UCIs such as pipelines, internet lines, and power cables are part of a complex network of critical infrastructure elements on the bottom of the seas.
A complex system can be analyzed by understanding how its components interact with each other by using network science~\cite{Milanovic18}. 
A network representation offers a common language to study different types of UCIs through graph theory, where a network is described by the system's components called nodes or vertices, and the direct interactions between them, called links or edges.
The structure and topology of such a complex network plays an essential role in a system's ability to
survive random failures or deliberate attacks.
In a network context, the system's ability to carry out
its basic functions even when some of its nodes and links may be missing is referred to as \textit{robustness}.
Additionally, a system is \textit{resilient} if it can adapt to component failures by changing its mode of operation, without losing its ability to function. 
In order to improve resilience, it is important to identify the most critical nodes of networked UCIs that are most likely to be failure points or vulnerable to attacks,
and
assess the consequences. %
Recent research on robustness and resilience of complex networks to failures and attacks include, respectively,~\cite{Barabasi2000} and~\cite{Kong10}.
Moreover, different models (e.g.,~\cite{Motter02,Tanizawa05}) have been proposed to capture the dynamics of cascading failures in systems characterized by some flow (e.g., information, natural gas, electric current) over a network. This allows us to understand the fraction of nodes that can be removed before global connectivity of the network is lost, how to stop a cascading failure, and how to enhance a system's dynamical robustness. 
As already anticipated in Section~\ref{sec:uci}, considering the underwater cable infrastructure, maritime choke points can be identified as critical points due to their high density of cables and maritime traffic. According to a recent analysis on security threats and consequences for the EU~\cite{Bueger22}, two key maritime bottlenecks are the Strait of Gibraltar and the passage between the Indian Ocean and the Mediterranean Sea via the Red Sea, respectively. The former, connecting the Mediterranean Sea and the Atlantic Ocean, is a dense area used for various maritime activities including submarine activities, with seven intercontinental cables passing through the strait. The latter represents the core connectivity to Asia where intercontinental cables pass through the Egyptian mainland adjacent to the Suez Canal to enhance the system's dynamical robustness.

Understanding and analyzing the interaction and interdependencies among UCIs is of utmost importance. Interdependency is a bidirectional relationship between two infrastructures through which the state of each infrastructure influences or is correlated to the state of the other. Types of interdependencies include the following~\cite{Rinaldi01}.
\begin{itemize}
    \item Physical interdependencies, which arise from physical links or connections among elements of the network. In this context, disruptions and perturbations in one component can propagate to other elements.
    \item Cyber interdependencies, which occur when the state of a component depends on data transmitted through the information infrastructure. Such interdependencies result from the increased use of computer-based information systems for monitoring and management activities (e.g., SCADA).
    \item Geographic interdependencies, which exist between two infrastructures when a local environmental event can provoke changes in both of them. This generally occurs when the components are in close spatial proximity, e.g., infrastructures that cross borders or that provide cross-border services,
    thus impacting the interests of different nations.
    \item Finally, logical interdependencies, which gather all interdependencies that are not physical, cyber or geographic, caused
    by, e.g., regulatory, legal, or policy constraints.
\end{itemize}
   
\subsection{Legal aspects}
The legal status of underwater pipes and cables varies based on the legal zone in which they are located, as determined by the United Nations Convention of the Law Of the Sea (UNCLOS). 
Within a country's territorial waters, which extend up to 12 nautical miles from the coastline, the country has full jurisdiction over the pipe/cable.
In the contiguous zone, which extends from 12 to 24 nautical miles from the coastline, states have specific law enforcement duties and obligations. 
Outside of these zones, particularly in the high seas (areas outside of national jurisdiction) as well as in the exclusive economic zones of states (i.e., up to 200 nautical miles
from a nation's
coastline), the legal status of pipes/cables and responsibility for their protection is currently defined as ``unclear'' and ``ambiguous''~\cite{Bueger22, unclos}.
Furthermore, underwater pipes may also be subject to 
additional regulations and laws depending on the activities they are associated with and their specific use. To make an example, oil and gas underwater pipes may be subject to additional environmental protection and safety-related regulations, while underwater pipes crossing national borders (or used for international trade) may be subject to additional international laws and agreements.

\section{Sensors and Contextual Information}
\label{sec:sensors_contextutal_information}
In this section, we describe
different types of sensors and technologies that are available and can be exploited for
UCIs surveillance.

\subsection{Coastal radars and AIS}
\label{subsec:coastal_radars}
Among the surveillance sensors commonly used for
MS, S-band or X-band
pulse radar sensors installed along the coastline represent a relevant and consistent source of information~\cite{VivoneBH15_53}. However, their coverage area and maximum range might be limited by line-of-sight propagation.
This limitation can be overcome by the installation and employment of long-range sensors such as 
High-Frequency Surface-Wave (HFSW) radars which have been considered 
for ship localization and tracking~\cite{Aureliano13,LiXZQLB16,BracaMGBH15_30,PonsforW10_18, MarescaBHG14_52}.
Initially introduced for ocean remote sensing, HFSW radars could dramatically increase MS coverage
by their ability to detect targets at Over-The-Horizon (OTH) distances.
In particular, multiple HFSW radars~\cite{BracaMGBH15_30,PonsforW10_18, MarescaBHG14_52}, 
combined with other data sources such as
AIS, satellite images, and contextual information, have the potential to provide continuous-time coverage of large sea areas at OTH distances~\cite{Fusion_AIS,GagSolMeyHlaBraFarWin:J20}. 

Besides conventional
and OTH coastal radars, AIS --- an anti-collision
broadcast system of transponders automatically exchanging ship traffic information for maritime 
safety --- definitely represents the major source of information by volume and granularity on surface vessel traffic.
According to the International Association of Marine Aids to Navigation and Lighthouse 
Authorities (IALA), the scope of AIS is ``to improve the maritime safety and efficiency of 
navigation, safety of life at sea and the protection of the marine environment'' 
~\cite{iala2003technical}.
In 2002, the International Maritime Organization (IMO) Safety of Life at Sea (SOLAS) convention~\cite{UN_solas} included a mandate that requested many commercial vessels to fit onboard AIS.
Specifically, IMO requires ships over 300 Gross Tonnage (GT), cargo vessels over 500 GT, all 
passenger ships, and all fishing vessels over 45 meters (in EU countries over 15 
meters~\cite{EC2011}) to be equipped with an AIS transponder onboard.
AIS messages can be exchanged through both satellite and terrestrial receivers~\cite{Space_1_AESM} and convey information about ship identifier, i.e., the Maritime Mobile Service Identity 
(MMSI), route (position, speed, course, and true heading), and other ship and voyage
information, including ship and cargo type, size, destination, 
and estimated time of arrival.

The analysis of AIS trajectories is used, among others, to pinpoint potentially illegal or illicit activity performed by vessels in specific areas of interest (see, e.g.,~\cite{Enrica18,Enrica21}).
A first filtering of AIS trajectories could be performed according to some selection criteria. In particular, considering a bounding area of $D_{\text{max}}$ kilometers around an UCI or a point of interest, e.g., Nord Stream explosion point, one could select all the AIS trajectories that have spent at least a given period of time $T_\text{min}$ inside the area. A further filtering to spot stationary or drifting ships could be achieved by selecting those trajectories whose average speed is lower than a maximum speed  $S_{\text{max}}$, or whose average rate of
manoeuvres per minute is higher than a predefined threshold. Non-kinematic information may be useful as well for AIS-based filtering. For example, a non-kinematic-based selection criteria may consist of excluding vessels whose small size would not allow them to represent a possible threat to underwater pipes (e.g., fishing vessels in very deep waters). A combination of kinematic-based and non-kinematic-based selection criteria can also be taken into account.%

In reality, AIS messages can be counterfeited and AIS transponders can be easily switched off, or vessels could navigate outside the coverage of coastal/satellite AIS receivers. For these and other reasons, AIS data is often complemented with data from other sensors or sources; for instance, %
historical geo- and time-referenced images provided by 
space-based sensors, e.g., SAR, multi-spectral (MSP), and hyper-spectral (HSP) sensors. 

\subsection{Satellite sensors: SAR, MSP and HSP}
\label{subsec:sat_sensors}
SAR is a high-resolution imaging system typically employed on board satellites (or aircrafts) that has become %
essential for wide-area monitoring to detect and track vessels at sea independently
of their compliance with the
SOLAS convention.
It is an active
remote sensing technology
based on the transmission of an electromagnetic (EM) microwave signal toward the Earth, 
and the reception and processing of the signals scattered by any natural or artificial features on the surface.
The use of EM microwave signals --- that undergo a weak scattering and absorption from the atmosphere --- makes the
monitoring capability of SAR systems
independent of sunlight illumination, thus allowing an effective sensing regardless of the weather conditions.
SAR systems typically operate in a monostatic geometry, where the receiving antenna is co-located with the transmitting one. In this configuration, only the energy reflected in the {\em backscattering} direction is collected by the system.
The received signals are then
properly processed to form a 2-dimensional (2D) image of the scene reflectivity in slant-range/azimuth coordinates. The features of the obtained image largely depend on both surface parameters, e.g., material composition, small- and large-scale roughness, and sensor parameters, e.g., viewing angle, frequency, polarization, and spatial resolution.

While active microwave remote sensing systems, such as SAR, offer a key support for gathering rather coarse information at multiple frequencies, polarizations, and viewing angles, MSP and HSP optical sensors can be exploited fruitfully to infer more accurate details in the spatial and spectral domains. 
MSP and HSP sensors cover the entire optical region --- between microwaves and X-rays ---
which refers to wavelengths ranging from 0.3 $\mu$m to 15 $\mu$m. Space-based MSP missions allow to collect imagery easy to interpret at a usually higher spatial resolution than SAR systems;
however,
MSP sensors are sensitive to cloud and sunlight conditions, and
they usually cover limited areas during each acquisition. 
On the other side, HSP satellite images are characterized by very high-spectral resolution and are suited to accurate classification, but they have low spatial resolution,
require a large computational burden, and are as well sensitive to cloud and sunlight conditions. 
For the interested reader, a comprehensive review of satellite sensors can be found in~\cite{Space_1_AESM}.

\subsection{UUVs equipped with acoustic sensors}
\label{ref:acoustic_sensors}

\begin{figure*}[!t]

	\centering
	\subfloat[Density map of the entire maritime traffic of the Baltic Sea.]{
		\includegraphics[width = .9\columnwidth]{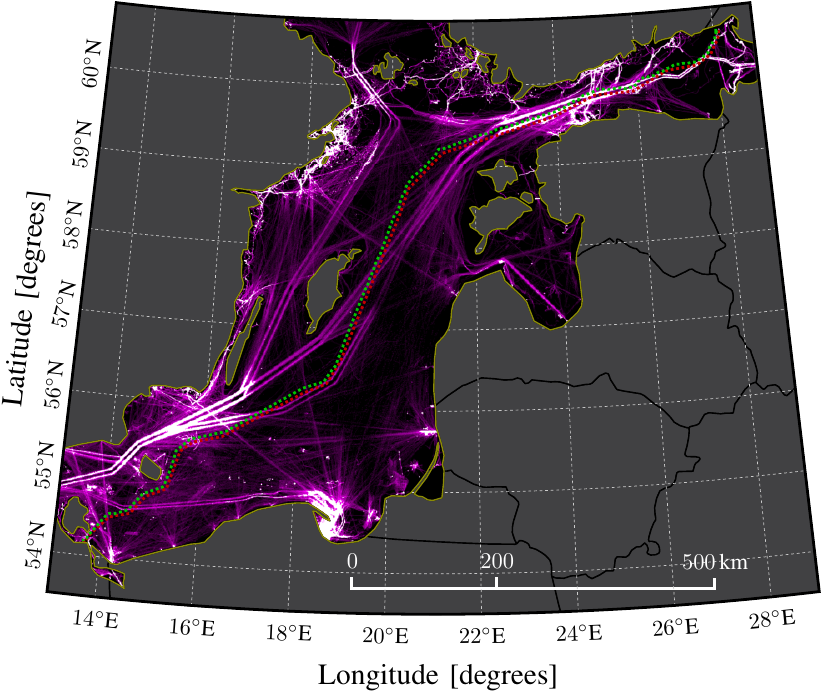}
		\label{fig:density_a} 
	}
	\centering
	\subfloat[Density map of the stationary areas of the Baltic Sea.]{
		\includegraphics[width = .9\columnwidth]{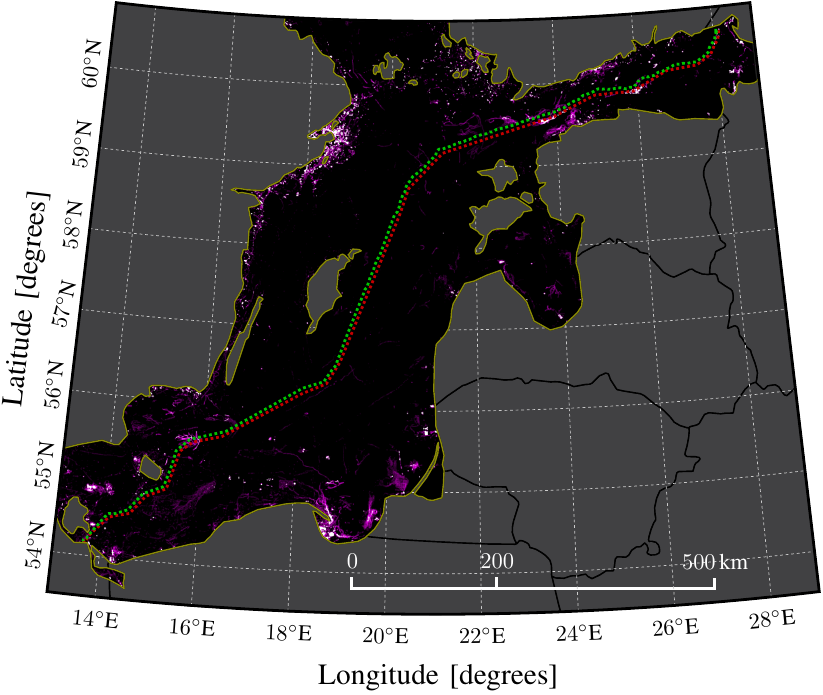}
		\label{fig:density_b}
	}
	
	\caption{Density maps of the  maritime traffic in the Baltic Sea built using AIS data from September 1st to September 15th, 2022. A higher brightness corresponds to a higher traffic density. Green and red dotted lines show the Nord Stream 1 and 2 gas pipelines, respectively.}
	\label{fig:density-map}
	
\end{figure*}

Unmanned Underwater Vehicles (UUVs) equipped with acoustic sensors are 
rapidly becoming the predominant platforms for undersea observation and 
monitoring. UUVs can be generally classified into Remotely Operated Vehicles (ROVs) and Autonomous Underwater Vehicles (AUVs)~\cite{Braca_ambiguity}. The former ones are 
tethered to ships or marine platforms and operated from above the water's 
surface. The tether allows operators to receive sensor data, e.g., sea-bottom images from cameras, and, if needed, guide remotely the ROVs almost in real-time; however, the operation range is limited by the tether's length.   
On the other side, AUVs are untethered and computer-controlled, with little or 
no operator interaction while performing their pre-programmed subsea mission; multiple AUVs can also cooperatively form an intelligent sensing network for the monitoring of large regions of interest~\cite{Ferri_coop, Brambilla_coop}.
Even though AUVs can be programmed to survey larger areas than ROVs, their 
operating range is
dependent on
the duration of their batteries. Furthermore, 
underwater communications, commonly exploiting the sound channel, are 
unreliable and characterized by limited bandwidth and range, thus limiting the 
ability of AUVs to effectively share sensor information in 
real time. 
Therefore, the choice among ROVs
and
AUVs is
dependent on the mission requirements as well as on maximum range, operating depth, time to cover a 
required distance,
and type and size of sensors they could bring on board, e.g.,
acoustic, 
magnetic, optical, and oceanographic.

Since optical and 
magnetic waves do not propagate well in seawater,  acoustic sonars, which employ sound waves to detect and  consecutively localize underwater objects, are 
nowadays the most common technology for undersea surveillance.
Passive sonars rely on the reception and  processing of acoustic 
information that is radiated by underwater noise sources
e.g., the noise produced by ships or submarines propellers; 
active sonars, instead, send an acoustic waveform and process the signals reflected by underwater objects.
Synthetic aperture sonars (SAS)~\cite{SAS_article} represent an established technology to collect 
high-resolution images of the seabed and underwater infrastructures. Similarly 
to SAR, a SAS continously transmits acoustic signals and
combines successive received pulses reflected by an object or a surface along a known track to create a 2D image of the illuminated area.

\subsection{Distributed acoustic sensing}
\label{subsec:dist_acoustic_sensing}
Distributed Acoustic Sensing (DAS) is an emerging technology which is commonly employed for the detection and analysis of seismic waves on the ocean bottom and for submarine structural characterization~\cite{Cheng2021,Trafford2022}. It is  enabled by fiber optic installed along underwater infrastructures, that continuously 
allows the monitoring in real time of underwater assets. While traditional 
monitoring systems rely on discrete sensors measuring at pre-fixed points, 
the fiber optic cable enables a continuous monitoring along a very long 
portion of the underwater infrastructure.  
Even though current commercial DAS systems allow a thorough monitoring along 
a maximum distance of 50 kilometers, recent studies have shown that persistent monitoring could be enabled up to a hundred kilometers. 
The most common DAS technologies are based on phase 
sensitive Optical Time Domain Reflectometer ($\phi$-OTDR) and coherent
OTDR (C-OTDR)~\cite{DAS_1}. A DAS interrogator unit 
generates a series of laser pulses,
sends them through the optical 
fiber cable, and collects the backscattering of the light along the length of 
the fiber. The analysis of the backscattered signal
by means of classification algorithms allows to detect and locate events such 
as leaks, intrusion activities, cable faults, or other anomalous events.

\subsection{Contextual Information}
\label{subsec:contextual_information}
Contextual information is generally intended as information that does not directly refer to the assets under surveillance, but to their surroundings. Contextual information adds to the operational picture the clarity that is needed to drive the actions to be taken. As such, contextual information is seldom conveyed by a single piece of information alone, but is rather derived from a mixture of experience, domain knowledge, and data artifacts. In the MS setting, examples of contextual information include geographic databases, such as the bathymetry and the displacement of critical infrastructures; geospatial information, such as meteorology and oceanography; intelligence reports, comprising human intelligence--HUMINT--and open-source intelligence--OSINT; reports on business ownership structures, sanctions, and criminal behavior of ship owners; and derived information from past/historical data, such as maritime Patterns-Of-Life (POLs).
To make a practical example, a vessel's trajectory 
could raise %
major concern %
if the vessel was previously involved in criminal activities, if only opaque ownership-related information of the vessel is available, or if there is evidence of the vessel deploying specialized equipment in proximity of sensitive infrastructures. Moreover, taking into account bathymetry is crucial, since the difficulty and risk in performing sabotage operations
are directly proportional to the sea depth. 
Furthermore, POLs can be considered as particular sets of behaviors and movements, e.g., waiting, navigating, or drifting, associated with specific entities, e.g., fishing vessels, cargo vessels, and oil tankers, over a defined period of time. In this context, density maps, built using historical AIS data in a given time interval and area, provide a preliminary insight on the most common POLs. This crucial information can be used for a preliminary classification of AIS trajectories. In fact, a selected AIS trajectory can be considered more or less suspect depending on whether its behavior seems compatible with the detected POLs in the given period of time  and area. Figure~\ref{fig:density_a} shows a density map of the entire maritime traffic in the Baltic Sea, built using AIS data collected from September 1st to September 15th, 2022. Purple patterns highlight the most common maritime routes in the considered region. The same data have been used to derive a density map of the stationary areas as shown in Fig.~\ref{fig:density_b}.%

\section{Seabed-to-Space Situational Awareness (S3A)}
\label{sec:seabed-to-space-situational-awareness}

The major challenge that operators and analysts face is identifying patterns emerging within very large datasets, e.g., AIS, SAR, optical, and multispectral data, when the goal is to anticipate possible future behaviors of suspicious assets and the related threats. In this context, information undoubtedly plays a crucial role, and artificial intelligence (AI) opens up unprecedented possibilities for surveillance systems to improve
MS, and in particular the resilience of UCIs. AI and Information Fusion (IF) can easily process immense volumes of information, fused from a variety of sources and generated from a very large number of monitoring assets on a day-to-day basis, thus enabling a potential future transition to an holistic perspective of S3A. 
The learned knowledge therefrom can be used as a valuable support to the cognitive processes (perception, comprehension, and projection) of analysts and operators to anticipate future behaviors and/or identify threats and critical situations that might endanger UCIs.
In the following, we will provide an overview of state-of-the-art Bayesian
IF and Multi-Target Tracking (MTT), anomaly detection, and automatic reasoning techniques, that might enable S3A and improve the
monitoring of UCIs.

\subsection{Bayesian IF and MTT}
\label{subsec:bayesian_inf_fus_MTT_MS}

The main objective of a multisensor MTT method is to sequentially estimate the number of targets together with their states, e.g., position, velocity, course, 
and heading, in a particular maritime area, by fusing measurements from multiple heterogeneous sensors. Each measurement is either a noisy observation of a 
target's kinematics, shapes, or other features, or a false alarm. 
In a Bayesian formulation, the MTT method amounts to estimating at each time (approximations of) the marginal posterior distributions of the detected targets' states, using all the measurements available up to the current time. MTT methods have to deal with various challenges, for example, the heterogeneity of the different information sources~\cite{Fusion_AIS,GagSolMeyHlaBraFarWin:J20},
asynchronicity, out-of-sequence measurements \cite{MilBraBryWil:C15},
latency,
and the measurement-origin uncertainty (MOU)~\cite{MeyKroWilLauHlaBraWin:J18}, i.e., the fact that it is unknown which target (if any) generated which measurement.
Existing MTT algorithms can be broadly classified as \textit{vector-type} algorithms, such as the joint probabilistic data association
filter~\cite{BarWilTia:B11} and the multiple hypothesis tracker~\cite{Rei:J79, ChongMR18}, and \textit{set-type} algorithms, such as the (cardinalized) probability 
hypothesis density filter~\cite{Mah:J03,Mah:J07}
and multi-Bernoulli filters~\cite{VoVoCan:J09}.
Vector-type algorithms represent the multitarget states and measurements by random vectors, whereas set-type algorithms represent them
by random finite sets.
Algorithms of both types have been developed and evaluated,
and several limitations have been noted~\cite{GagSolMeyHlaBraFarWin:J20}. First, the fusion of 
heterogeneous information sources is not straightforward. Second, they do not adapt to time-varying model parameters. And third, their complexity usually does 
not scale well in relevant system parameters, e.g., the number of sensors.

An emerging approach to MTT and IF --- one with flexibility, low complexity, and useful scalability --- is based on a factor graph and the 
\emph{sum-product algorithm} (SPA)~\cite{MeyKroWilLauHlaBraWin:J18}.  
First, a factor graph representing the statistical model of the MTT problem is derived; then, the SPA is used to  solve efficiently the MOU problem and obtain a principled and intuitive approximation 
of the Bayesian inference needed for target detection and estimation.
A major advantage of the SPA is its ability to 
exploit conditional independence properties of random variables for a drastic reduction of complexity; thereby, SPA-based MTT algorithms can achieve an 
attractive performance--complexity compromise, making them suitable for large-scale tracking 
scenarios involving a large number of targets, sensors, and measurements, and allowing their use on resource-limited devices. Generally, the complexity of an 
SPA-based MTT algorithm scales
quadratically in the number of targets, linearly in the number of sensors, and linearly in the number of measurements per 
sensor, and outperforms previously proposed methods in terms of accuracy (see~\cite{MeyKroWilLauHlaBraWin:J18} and references therein).%

SPA-based MTT algorithms can be easily extended to automatically estimate unknown and time-varying parameters~\cite{SolMeyBraHla:J19}, such as detection probabilities of sensors, incorporate multiple dynamic models, and in particular fuse heterogeneous data~\cite{Fusion_AIS}, e.g., from terrestrial radars, SAR, optical sensors, and AIS. 
In the context of S3A and in particular UCIs monitoring, the fusion of satellite data, e.g., SAR, with AIS potentially allows the identification of ships, which switch off the AIS during anomalous activities or sabotage operations. 
The fusion of this information is often difficult due to the asynchronicity and sparsity of AIS messages, and the non-trivial association between messages and targets. 
Indeed, although each AIS message usually includes a unique
identifier--the MMSI--, this may be absent,
incorrectly received, or
observed for the first time, in which case no prior information is available on the target--message association. 
The SPA-based MTT method can be efficiently extended to fuse AIS messages and measurements obtained from SAR, optical images, or other sensors, and to 
identify (or label) each detected ship by means of the MMSI.

\subsection{Anomaly detection}
\label{subsec:anomaly_detection}

Increasing automation through a large number of advanced data-driven methods and techniques for maritime anomaly detection has enabled the system and the operator to spot complex situations by correlating various events from all surveillance sensors and classify them into important incidents~\cite{Laxhammar2008,Sidibe2017,NextGen}. 
\begin{figure*}[!t]
\centering%
	\subfloat[Full Sentinel-1 SAR image over the Baltic Sea.]{
		\includegraphics[width = .9\textwidth]{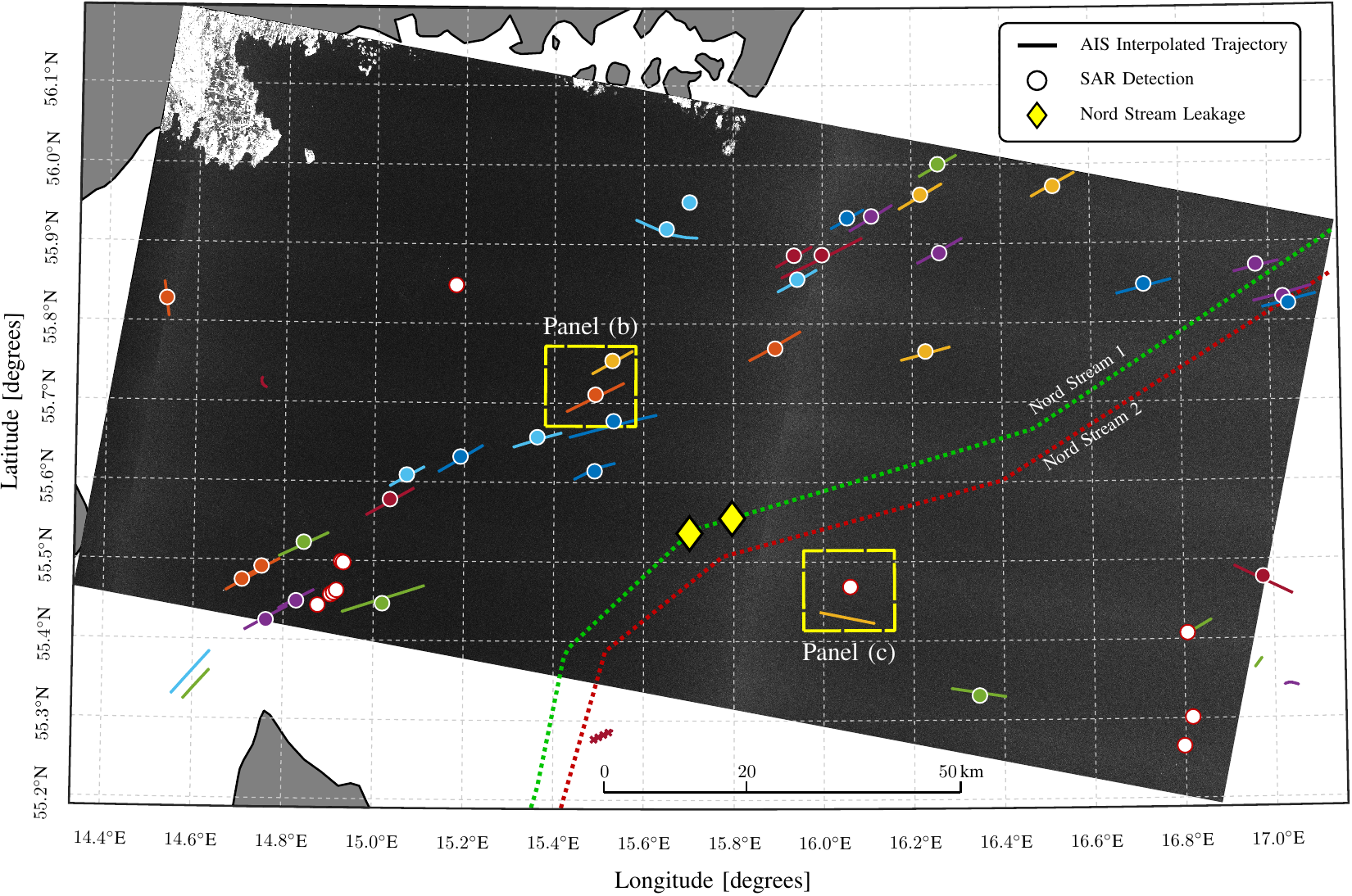}
		\label{fig:full-sar-image} 
	} \\[0mm]
	\subfloat[Detail of the SAR image showing three associated detections.]{
		\includegraphics[width = .9\columnwidth]{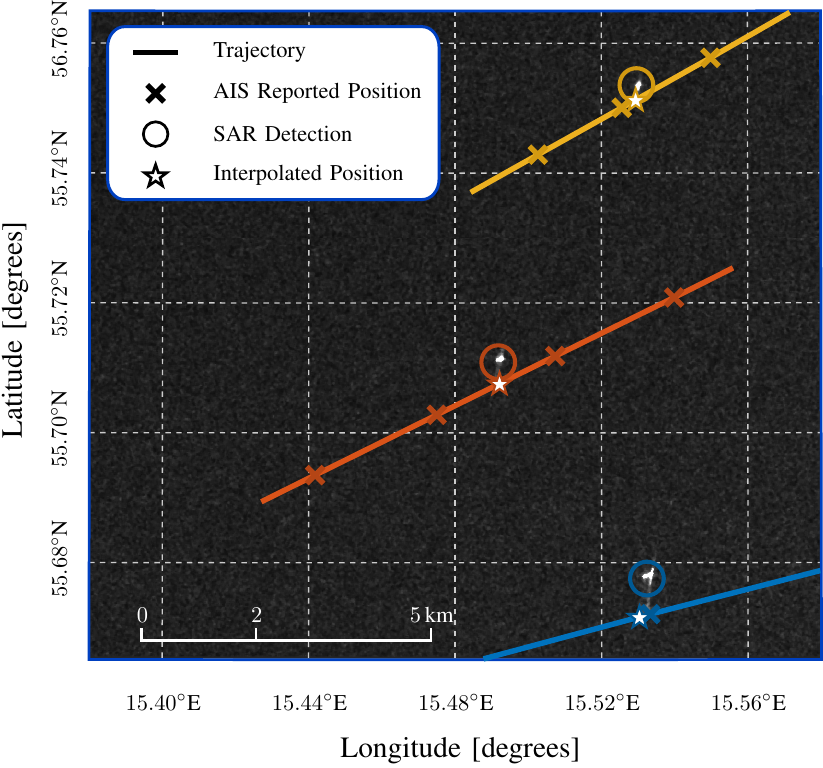}
		\label{fig:sar-detail-1} 
	}
	\subfloat[Detail of the SAR image showing one not-associated detection.]{
		\includegraphics[width = .9\columnwidth]{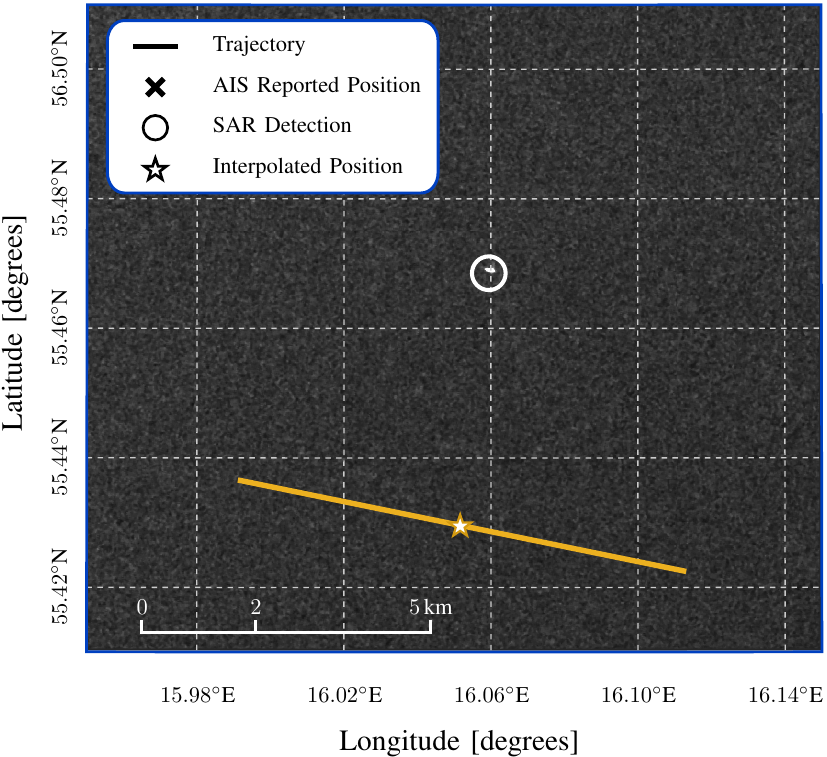}
		\label{fig:sar-detail-2}
	}

	\caption{Nord Stream use case. Sentinel-1 SAR image acquired over the Baltic Sea some days before the
    explosions.
	Colored solid lines are vessels' trajectories obtained by interpolating the AIS positions provided by the vessels themselves.
	Circles are ships as detected within the SAR image: the
	color of the circle identifies the detection as coming from a vessel whose AIS trajectory has the same color; a white circle indicates that the detection is not associated with any AIS trajectory.
	Panels \protect\subref{fig:sar-detail-1} and \protect\subref{fig:sar-detail-2} present two details of the full SAR image showing examples of associated detections and not-associated detections, respectively.
	Here, the crosses depict AIS reported positions, and the stars are the interpolated positions of the vessels at the time of the image acquisition.}
	\label{fig:sar-detection}
	
\end{figure*}
An anomaly in the maritime domain can be described as a behavior that is not ``normal" or, more specifically, not expected to occur during regular operations~\cite{Laxhammar2008}, and it can refer to a sudden change in vessel kinematic behavior (such as unusual speed or location), deviation from standard sea lanes, unexpected AIS activity, unexpected port arrivals, close approach, and zone entry~\cite{Lane10}. 
In particular, zone entry anomalies involve ships entering in protect environmental areas, or approaching military installations or UCIs.

\begin{figure*}[!t]

	\centering
	\includegraphics[width = .9\textwidth]{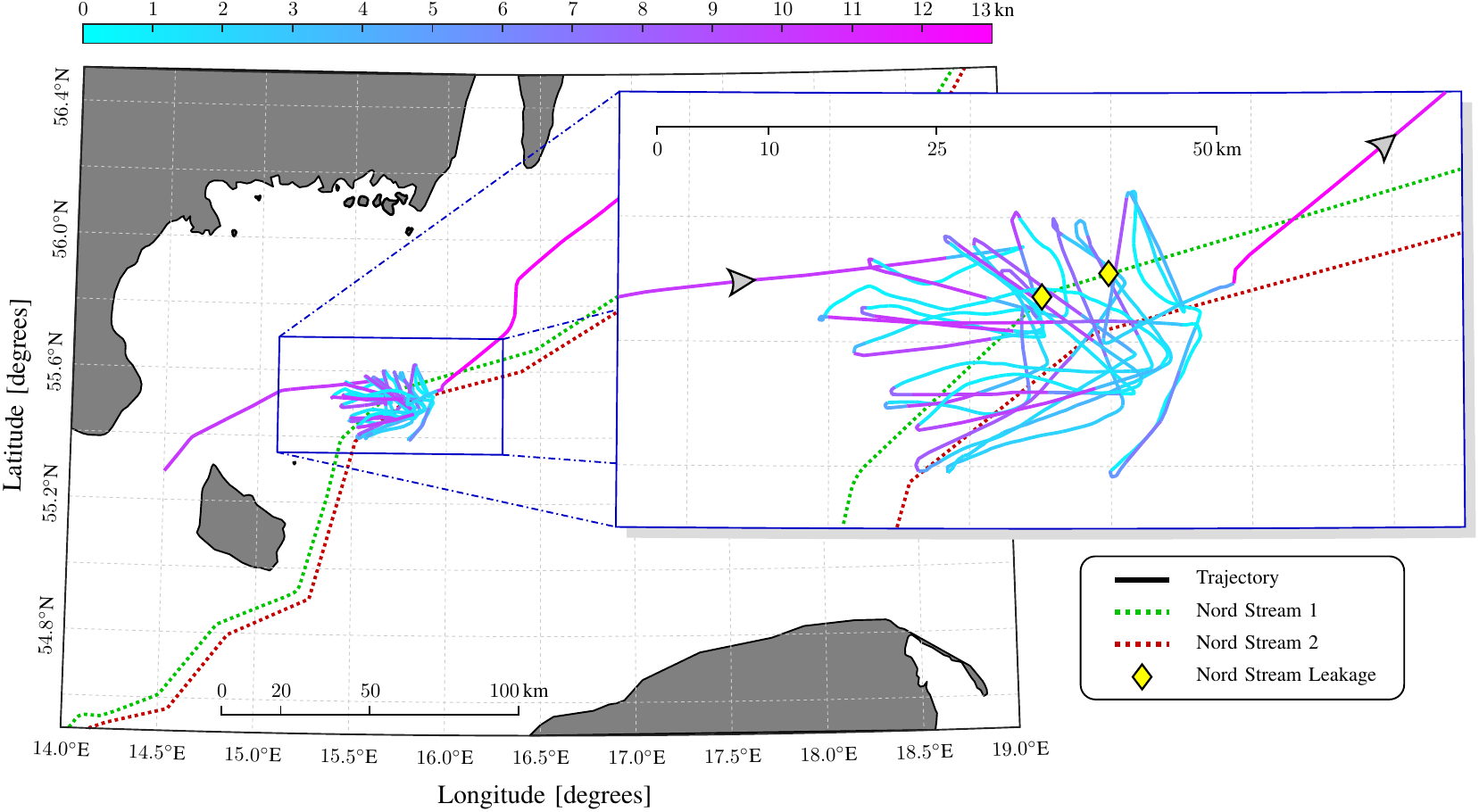}

    \caption{Nord Stream use case. Full trajectory of the vessel of interest a few days before the
    explosions.
	The trajectory spans several days and its color (cyan to magenta) is related to the vessel's speed (slower to faster); grey arrows indicate the direction of the vessel at the beginning and at the end of the trajectory.
	The locations of the gas leakage are marked by yellow diamonds.}
	\label{fig:full-trajectory}
	
\end{figure*}

Most anomaly detectors require learning an underlying model representing the normal behavior by using the available historical data. Based on the learned normalcy model, detectors can decide whether new data can be classified as normal or anomalous behavior. According to the available proposals and studies~\cite{Sidibe2017,Riveiro2018}, the data-driven methodologies for maritime anomaly detection can be divided into \emph{machine learning} and \emph{stochastic} approaches.
Machine learning techniques are able to identify patterns emerging within huge amounts of maritime data, fused from various uncertain sources and generated from monitoring thousands of vessels a day, so as to act proactively and minimize the impact of possible threats. The general aim of such an approach includes frequent pattern discovery, trajectory pattern clustering in a multidimensional feature space, trajectory classification, forecasting, and anomalies/outliers detection. The machine learning framework mainly comprises the following unsupervised and supervised methodologies. \emph{i)} Distance-based clustering methods are mainly based on the \emph{nearest neighbor} algorithm and implement a well-defined distance metric; the greater the distance of the object to its neighbor, the more likely it is to be an outlier. \emph{ii)} Density-based clustering methods identify distinctive groups/clusters in the data based on the idea that a cluster in a data space is a contiguous region of high point density, separated from other such clusters by contiguous regions of low point density. In particular, the Density-Based Spatial Clustering of Applications with Noise (DBSCAN) algorithm and its variations have become very popular for their convenient properties: these methods do not require specifying the number of clusters and have the ability to derive arbitrarily shaped clusters and identify outliers~\cite{Ester1996}. \emph{iii)} Classification methods require the construction of a classifier, that is, a function that assigns a class label to instances described by a set of attributes. Using supervised learning approaches, trajectories or segments of a trajectory can be classified into some categories, which can be motions, human activities, or transportation modes. The main classification methods include neural networks, support vector machines, and decision trees. 
Stochastic approaches to maritime anomaly detection fit a statistical model representing normal vessel behavior to the given historical vessel movement data, and then apply a statistical inference test to determine whether a new vessel observation belongs to this model or not. Observations that have a low probability of being generated from the learned model, based on the applied test statistic, are declared as anomalous behaviors. Bayesian networks, Gaussian processes, and the Gaussian mixture model represent the major methods within the stochastic approach.

\begin{figure*}[!t]

	\centering
	\subfloat[Extract of the trajectory also reporting the vessel's location as detected \\ in a SAR image acquired at time $T_{\text{A}}$.]{
		\includegraphics[width = .9\columnwidth]{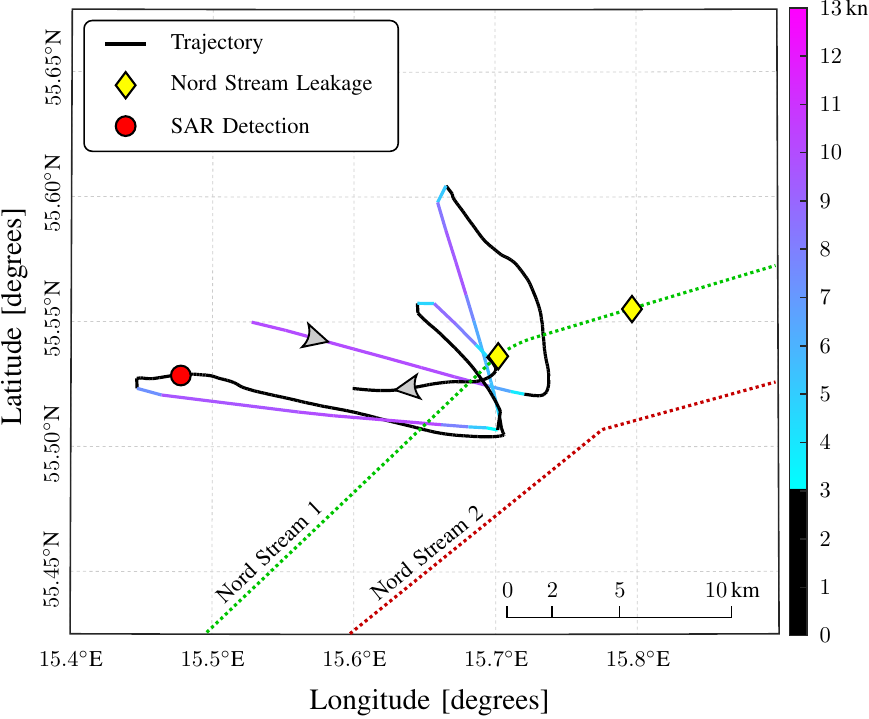}
		\label{fig:extract-trajectory-a} 
	}
	\centering
	\subfloat[Extract of the trajectory also reporting the vessel's location as detected \\ in a SAR image acquired at time $T_{\text{B}}$.]{
		\includegraphics[width = .9\columnwidth]{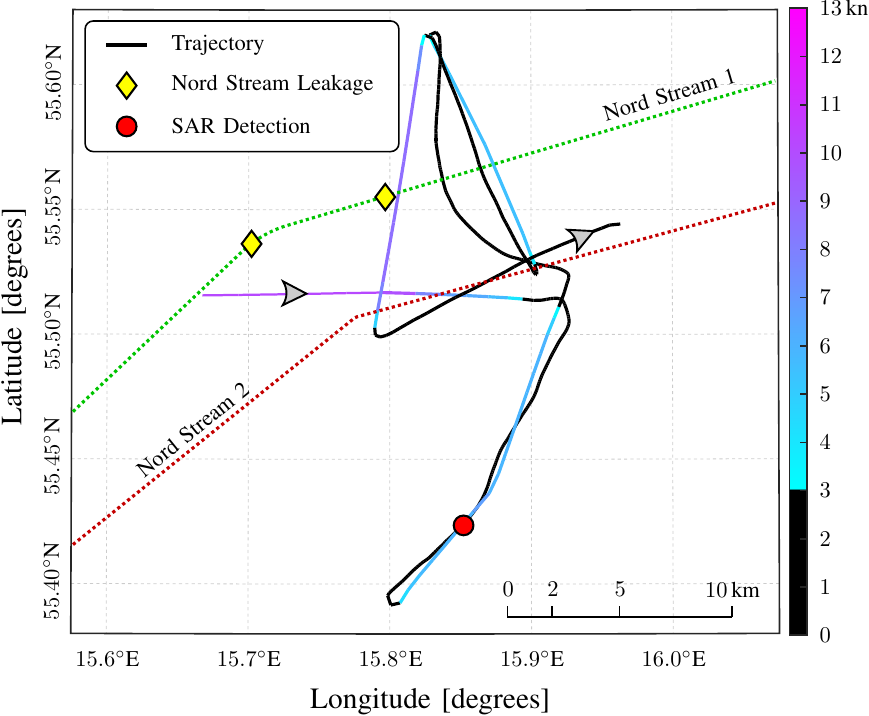}
		\label{fig:extract-trajectory-b}
	} \\[0mm]
	\centering
	\subfloat[Extract of the trajectory.]{
		\includegraphics[width = .9\columnwidth]{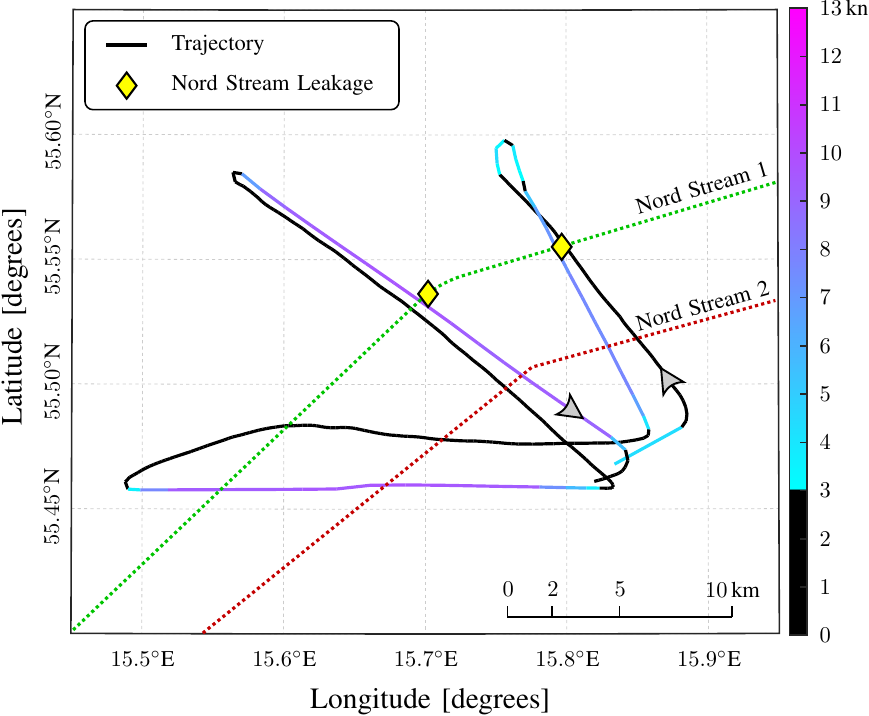}
		\label{fig:extract-trajectory-c} 
	}	
	\centering
	\subfloat[Extract of the trajectory.]{
		\includegraphics[width = .9\columnwidth]{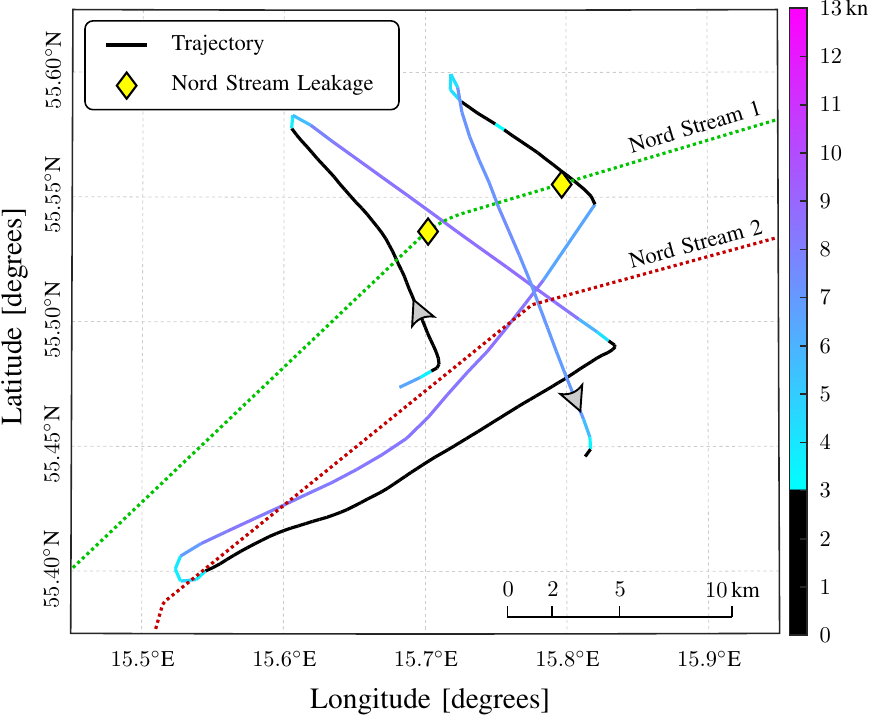}
		\label{fig:extract-trajectory-d} 
	}
	
	\caption{Nord Stream use case. Extracts of the full trajectory of the vessel of interest reported in Fig.~\ref{fig:full-trajectory}.
	Each panel reports a portion of the full trajectory that spans several hours.
	The colors (cyan to magenta) are related to the vessel's speed (slower to faster); speeds below 3 knots are all reported in black.
	Grey arrows indicate the direction of the vessel at the beginning and at the end of the trajectory, and yellow diamonds mark the locations of the gas leakage.
	A red dot (in panels \protect\subref{fig:extract-trajectory-a} and \protect\subref{fig:extract-trajectory-b}) reports the location of the vessel as detected in a SAR image; additional details on the SAR detections are presented in Fig.~\ref{fig:sar-image}.}
	\label{fig:extract-trajectory}
	
\end{figure*}

Current research on maritime anomaly detection considers the aforementioned types of anomalies.	The deviation from a standard route is the most prominent anomaly type that research addresses by extracting frequently traveled sea routes from historical AIS data, e.g., via clustering. Unknown AIS tracks can then be compared in order to investigate whether they are similar enough to the extracted routes, or in the case of clustering, belong to one of the identified route clusters. These approaches work very well in areas where many ships take similar routes~\cite{Ristic2008,Wang2014,Zhen2017}. A different approach~\cite{Enrica18,Ever_given} faces the same anomaly within a stochastic framework by combining the available context data with a parametric model of the vessel's kinematic behavior and running a hypothesis testing procedure to make decisions on the existence of anomalous deviations relying upon the available measurements (e.g., AIS, radar, SAR). 
Further stochastic strategies 
address the joint problem of sequential anomaly detection and tracking of a target subject to switching anomalous deviations in a Bayesian framework~\cite{Forti2018,Radarconf20,Forti2022}.
Other approaches particularly consider deviations from standard routes in the presence of unexpected AIS activity. Indeed, AIS tracks of ships are often characterized by large blind spots, and this may be due to unintentional technical problems, radio interference, attenuation, or actual equipment manipulation, such as intentionally turning off AIS transceivers~\cite{Mazzarella2017,Enrica18,Tserpes}. Moreover, AIS signals can be easily spoofed by external attackers or the crew itself willing to obscure their locations~\cite{Katsilieris,Enrica21}. The AIS intentional interruption or spoofing could indicate a will to hide some illegal activities, such as smuggling on coast or with other ships, or entry in unauthorized areas.
Zone entry as an anomaly type is considered only marginally~\cite{Lane10,Vespe}. Restricted zones with entry ban are learned implicitly as part of the general shipping routes and trajectories~\cite{Vespe}, whereas more elaborate methods such as predicting whether a zone entry is likely to occur soon are proposed in~\cite{Lane10}.
Anomalous port arrival is taken into account in~\cite{Zor} looking at ferries that run regular routes according to a fixed schedule, while close approach anomalies are investigated in~\cite{Lane10,Kowalska,Mascaro2014}.

\subsection{Reasoning for situation awareness}
Besides the data-driven approaches described earlier, there exist situations where higher-level reasoning needs to be considered.
This can be useful for aligning  highly heterogeneous information sources, which range from HUMINT and OSINT descriptions of the vessels' current and projected behavior, through contextual information to MTT tracks and anomaly reports.
This is in line with the Activity-Based Intelligence (ABI)~\cite{Llinas2014Foundational, Levchuk2009Adversarial} paradigm
that has been in use since the war in Afghanistan and
has brought a new vision of intelligence
pushing forward the development of multi-intelligence (multi-INT) capabilities, which aim
at considering in an exhaustive way all sources of information.
In order to make sense of this kind of data, each source needs to be corrected to account for its reliability and possibly contextual information.
In cases where different
sources provide reports on the same target property,
these reports can
be aligned
in accordance
to an appropriate
correction model;
if the reports refer to different properties, instead, they are used as inputs to a reasoning system that
verifies their consistency. For example, consider source A which reports a particular vessel as a tanker, and a source B which reports it to be a cargo vessel. Given that the reports are concerned with the same property with different degrees of semantic granularity, this information can be readily aligned. On the other hand, a source may report on the vessel type, and the other on its speed. In this case, the properties can be aligned using a reasoning system verifying the consistency of the speed given the vessel type. This can highlight conflict between the sources which can further lead to conclusions about spoofing or another anomaly, depending on the exact scenario.

Tackling hybrid threats is particularly challenging as it is necessary to predict an event, that, above all, is rare and can be essentially considered a black-swan event. As such, it is
unlikely to provide
an analysis based on data and machine learning only; in these cases
it is possible to leverage expert information using rule-based systems. 
One such approach, which allows a degree of semi-causal reasoning, involves using valuation-based systems with the Dempster-Shafer theory~\cite{Shafer.Mathematical.1976} (also known as the theory of belief functions). Such systems are also known as evidential networks. The information provided by the myriad of sources described earlier is corrected and aligned with a common vocabulary using a mechanism such as contextual belief correction~\cite{Mercier.Belief.2012}, behavior-based correction (BBC)~\cite{Pichon2016PropositionMechanisms}, or its context-aware extension~\cite{KOWALSKI202329}. A set of rules elicited from the experts is used to construct a valuation network which is defined by a set of variables (some of which may be directly mapped to the observations provided by the sources
whereas the others are inferred) and the relations between them~\cite{Shenoy1994UsingSystems}. %

Possible uses of such rule-based approaches range from trivial to significantly more complex ones. A trivial illustration could be reasoning about a vessel's inconsistent AIS status. For example, a trajectory classifier %
can be considered one source, and the vessel's AIS navigational status %
the other. If the navigational status is inconsistent with the type of trajectory provided by the classifier, the AIS information reported by the vessel
can be considered inconsistent.
This by itself is unlikely to mean that the vessel is involved in illegal activity, but
it may be an indication of anomaly which should lead to further investigation. 
More complex evidential networks have also been proposed in this context to
assess the vessel's intent, likelihood of criminal activity, or the overall threat level posed. 
One particular strength of the belief-theoretic approach provided by the evidential networks is that
the rules
are considered to be partially uncertain, thus
the reasoning can still be performed with missing data. This can be leveraged to identify which sources of information should be queried in order to improve the quality of the inference results.\
Finally, the transparent-by-design property of
expert systems facilitates the implementation of 
explanation abilities; these
describe the relative contribution of each source to the decision made,
the overall conflict and uncertainty in the answer obtained, as well as the impact of various uncertainties in data~\cite{Kowalski_2022}. 

Furthermore, for a better understanding of overall situation awareness which includes multiple entities interconnected by a set of relations, we can leverage conceptual graphs or knowledge graphs. Recent work on these has explored embedding uncertainty in such graphs and establishing links with belief functions theory and evidential networks in order to allow uncertain reasoning about multiple entities~\cite{9841301}.
In the context of hybrid threats, this is particularly relevant to synchronised threats originating simultaneously from several vessels coordinating their activity.  

\begin{figure}[!t]

	\centering
	\subfloat[Detail of the Sentinel-1 SAR image acquired at time $T_{\text{A}}$.]{
		\includegraphics[width = .9\columnwidth]{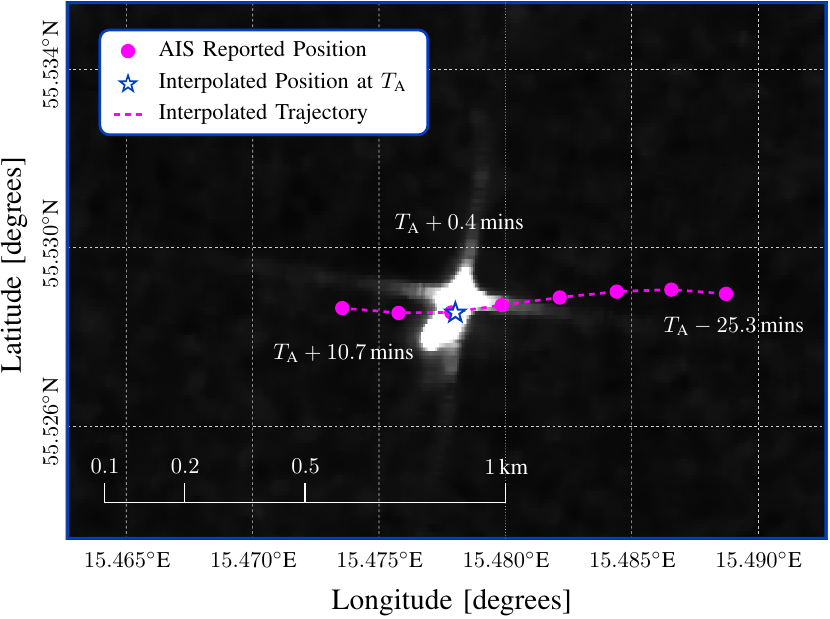}
		\label{fig:sar-image-a} 
	}
	
	\centering
	\subfloat[Detail of the Sentinel-1 SAR image acquired at time $T_{\text{B}}$.]{
		\includegraphics[width = .9\columnwidth]{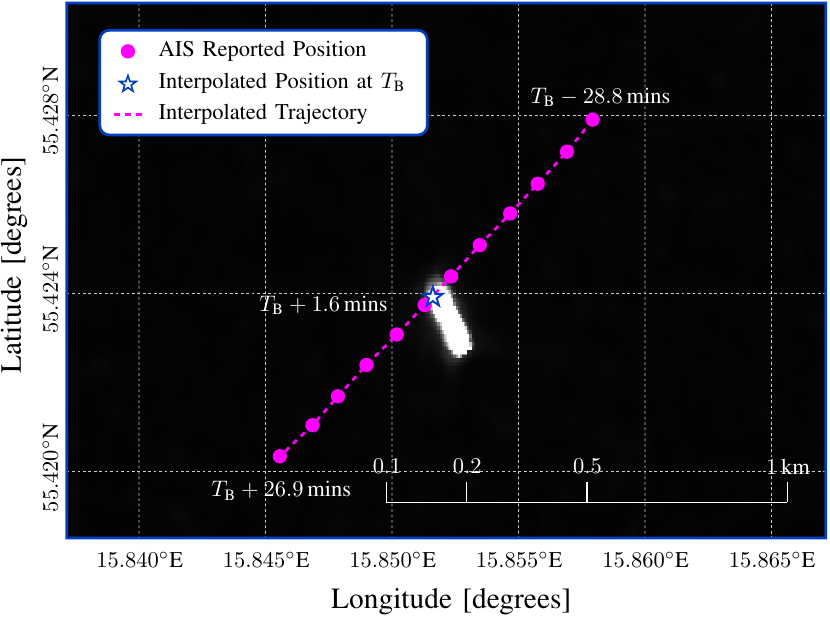}
		\label{fig:sar-image-b} 
	}
	
	\caption{Nord Stream use case. Details of two Sentinel-1 SAR images showing the vessel of interest.
	Panel \protect\subref{fig:sar-image-a} shows the SAR image of the vessel located as marked in Fig.~\ref{fig:extract-trajectory-a}; likewise, panel \protect\subref{fig:sar-image-b} shows the SAR image of the vessel located as marked in Fig.~\ref{fig:extract-trajectory-b}.
	$T_{\text{A}}$ and $T_{\text{B}}$ are the acquisition times of the SAR images, respectively, in panel \protect\subref{fig:sar-image-a} and panel \protect\subref{fig:sar-image-b}.
	Magenta dots represent the vessel's locations as reported by the AIS, while
	the blue stars indicate the vessel's
	interpolated locations
	at time $T_{\text{A}}$
	and time $T_{\text{B}}$.
	Dashed lines reproduce the vessel's trajectories obtained by interpolating the AIS reported locations.}
	\label{fig:sar-image}
	
\end{figure}

\section{Case Studies}
\label{sec:use_cases}
This section reports analyses on the Nord Stream and Shetland Islands accidents,
as well as on the anomalous behavior of a
large vessel in the Adriatic Sea,
and how surveillance sensors data, in particular AIS reported information and SAR images, might be beneficial for UCIs monitoring.

\begin{figure*}[!t]

	\centering
	\includegraphics[width = .9\textwidth]{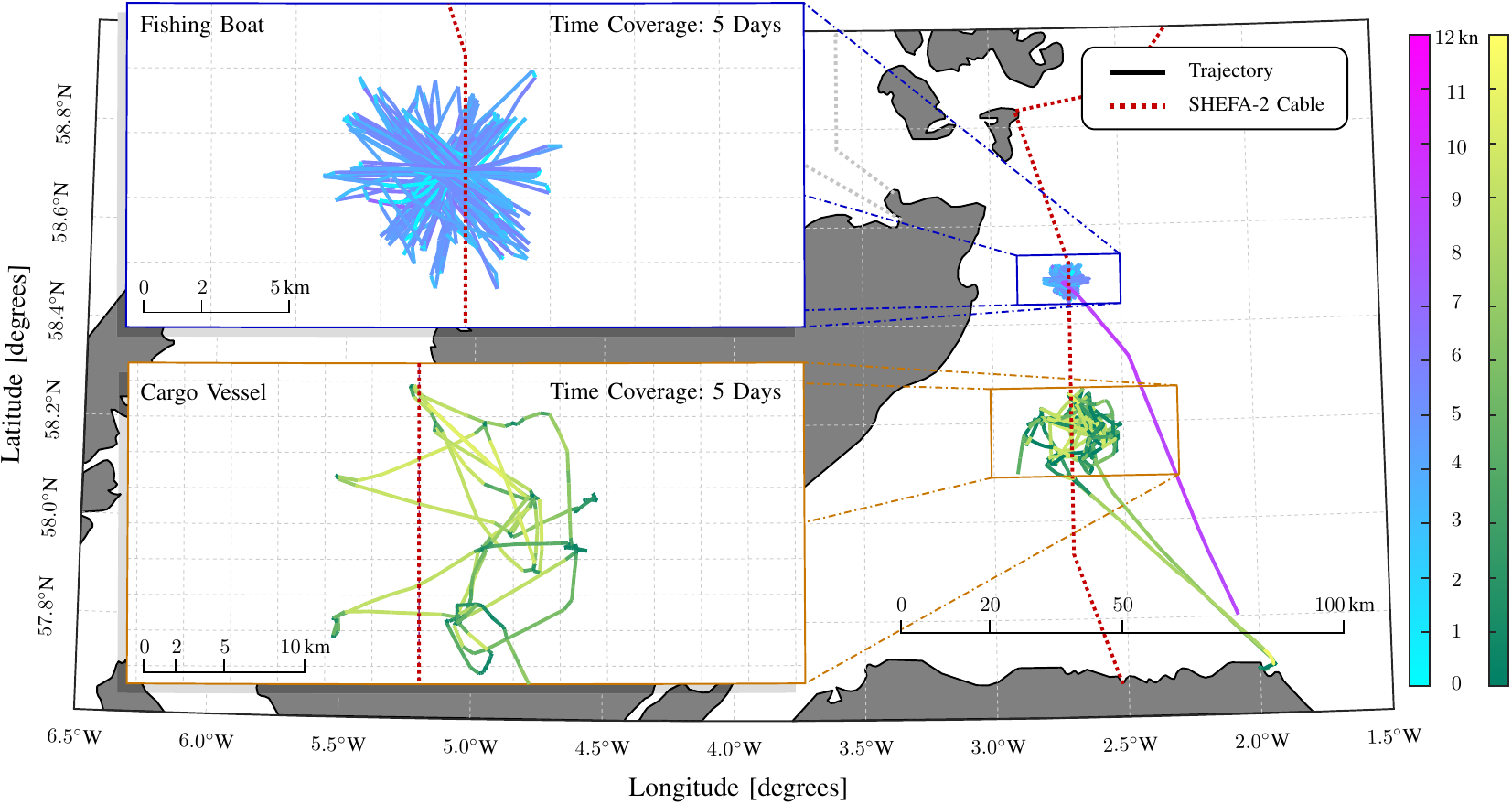}

	\caption{Shetland Islands use case. The main panel shows the full trajectories of a fishing vessel and a cargo vessel in the region of interest since few weeks before the SHEFA-2 cable
    cutoff.
	The inserts show details of the trajectories spanning 5 days.
	The color of the trajectories (cyan to magenta, and green to yellow) is related to the vessels' speed (slower to faster).
	}
	\label{fig:full-trajectory-shetland}
	
\end{figure*}

\subsection{Nord Stream}
\label{subsec:nord-stream}
Figure~\ref{fig:sar-detection} reports a Sentinel-1 SAR image acquired over the Baltic Sea some days before the explosions that affected the Nord Stream gas pipeline~\cite{nord-stream-press}. 
The circles indicate detected ships, while solid lines represent vessels' trajectories
that cover a time interval from 10 minutes before to 10 minutes after the SAR image acquisition time.
These trajectories are obtained by interpolating the AIS positions reported by the ships themselves, and filling gaps in the data of at most 6 hours.
The interpolation requires the availability of AIS positions both before and after the image acquisition time; when this information is not available, e.g., during (quasi) real-time operations, the vessel's position at the time of the image acquisition can be inferred from the available past data~\cite{MilBraBryWil:J16}.
The combined use of AIS data and satellite images (SAR, optical, MSP, and HPS) allows giving an identity to a picture of a vessel that could complement AIS for
MS and that, otherwise, would remain unknown.
On the other hand, the unavailability of AIS data for a detected ship within a satellite image can be caused by incorrect or lost data, or might highlight an anomalous behavior that requires further investigation.

The association between AIS trajectories and SAR detections in Fig.~\ref{fig:full-sar-image} is achieved by solving a specific assignment problem (clearly, other solutions are available in the MTT literature, discussed in the previous section). The assignment \textit{cost} between any SAR detection--AIS trajectory pair is the relative distance between the location of the SAR detection and the interpolated position of the vessel at the time of the image acquisition if such a distance is below 3 km, and it is assumed infinite otherwise.
The association is then represented
by the color of the circle that matches the color of the AIS trajectory which it is associated with; white circles represent SAR detections that are not associated with any AIS trajectory.
We observe that the majority of the vessels detected in Fig.~\ref{fig:full-sar-image} are associated with an AIS trajectory.
Figures~\ref{fig:sar-detail-1} and~\ref{fig:sar-detail-2} present two examples of associated detections and not-associated detections, respectively.
The first panel shows three detected ships, each within a colored circle.
The color of the circle matches that of the AIS trajectory, obtained as the interpolation of the AIS reported positions represented by the crosses.
The stars indicate the interpolated positions of the vessels at the time of the image acquisition; the offset between the shape of a vessel as detected by the SAR and its interpolated AIS position is due to the unknown Doppler frequency generated by the motion of the vessel itself, and it is thus related to its velocity~\cite{GraRenMoc:J19}.
Figure~\ref{fig:sar-detail-2}, instead, shows a single detected ship --- enclosed in the white circle --- which is not associated with any AIS trajectory; the closest available AIS interpolated position, indeed, is more than 3 km away.
This AIS trajectory, however, presents a relevant characteristic, that is,
a gap of several hours (yet less than 6 hours) in the AIS data before and after the acquisition of the SAR image.
On one hand, this could suggest that the detected object and the AIS trajectory depicted in Fig.~\ref{fig:sar-detail-2} do refer to the same vessel; on the other hand, the gap in the AIS data and the detection of the ship in a position that deviates from a linear path, might be an indication of an anomalous behavior.

From the analysis of the AIS data, the behavior of another vessel appears significant.
Figure~\ref{fig:full-trajectory} shows its full trajectory --- over several days ---
before the Nord Stream
accident.
The colors, from cyan to magenta, reflect the speed of the vessel.
Extracts of the trajectory, each spanning several hours, are reported in Fig.~\ref{fig:extract-trajectory}.
Speeds below 3 knots are reported in black because large vessels, such as the one considered, are hard to be maneuvered at such low speeds and tend to drift.
These portions of the trajectory may suggest that the ship is following a \textit{search} path, that is, it is maneuvering to approach the pipeline, and moving away from it while drifting.
Note that such behavior might also be compatible with other scenarios as, for example, loitering while waiting for orders.
Nevertheless, it is worth mentioning that the
region where the Nord Stream
accident occurred is not designed as a stationary area.

The presence of the vessel in the area is corroborated by its detection in two Sentinel-1 SAR images; Figs.~\ref{fig:extract-trajectory-a} and~\ref{fig:extract-trajectory-b} report the locations of the vessel as detected in the SAR images, while Figs.~\ref{fig:sar-image-a} and~\ref{fig:sar-image-b} show details of the mentioned \text{Sentinel-1} SAR images.
Both the SAR images are acquired while the vessel is drifting: this is confirmed by the vessel's speeds reported in Figs.~\ref{fig:extract-trajectory-a} and~\ref{fig:extract-trajectory-b} at the time of the acquisition (below 3 knots) and by the orientation of the vessel as reported in Fig.~\ref{fig:sar-image}.
Finally, note that the slight offset between the vessel's trajectory (dashed line) and the shape of the ship as acquired by the SAR, particularly evident in Fig.~\ref{fig:sar-image-b}, is compatible with the vessel's AIS reference point.

Beyond the aforementioned kinematic characteristics, two important remarks are useful to assess the potential implication of the vessel in the Nord Stream explosions. 
The former is related to the vessel's ownership risk assessment
that is considered very high,
thus not facilitating the identification of beneficial owners.
The second remark is related to the operations which can be conducted by the vessel: according to several navy officers, consulted by the authors, the vessel would have been perfectly able to support and coordinate a sabotage operation using specific instrumentation, such as a ROV. 

As discussed in the previous sections, all the aforementioned information needs to be ingested and automatically processed to asses an overall risk associated with a suspicious ship.  

\subsection{Shetland Islands}
\label{subsec:shetland}

\begin{figure*}[!t]

	\centering
	\includegraphics[width = .9\textwidth]{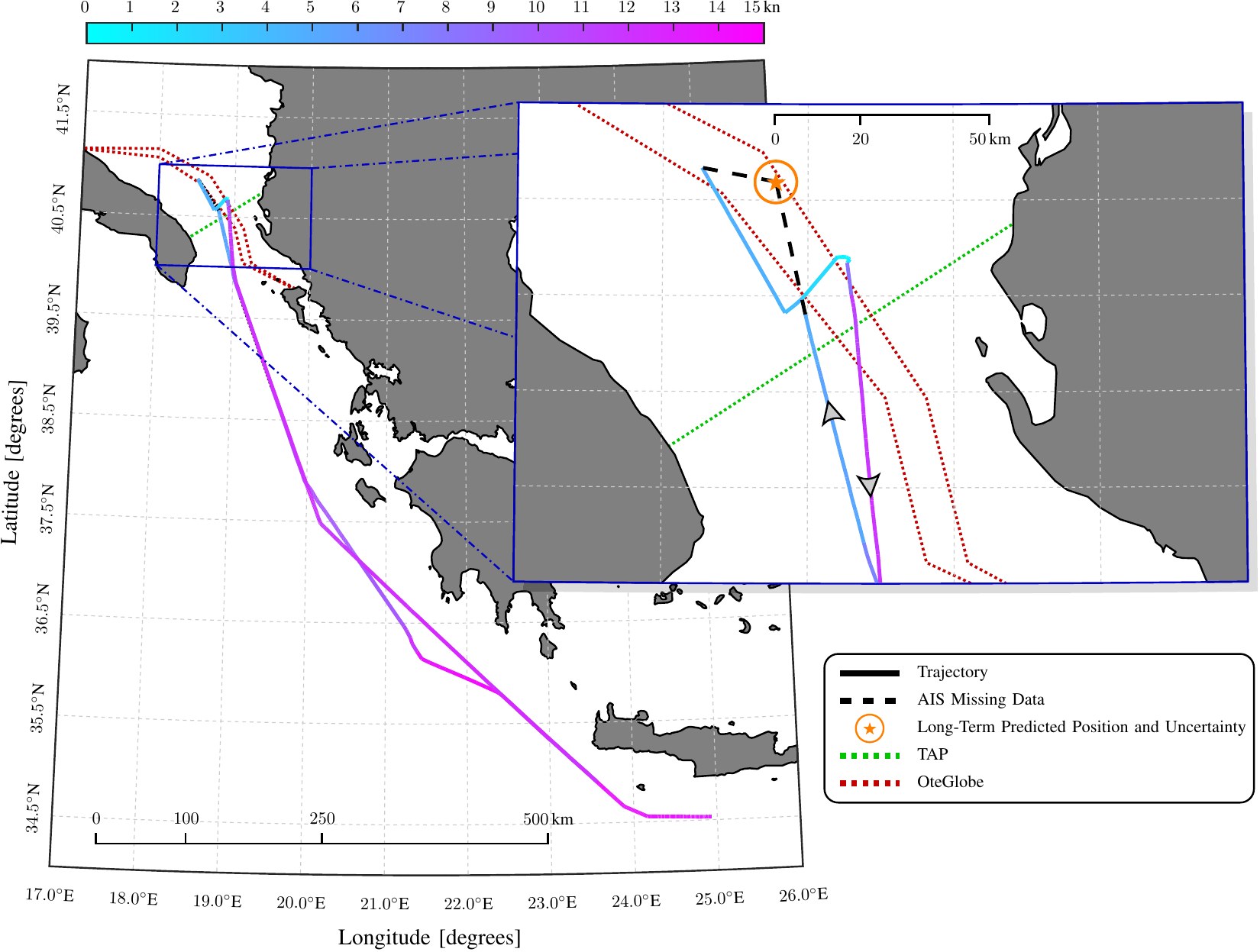}

	\caption{Adriatic Sea use case. Full trajectory of the
    vessel of interest from December 19th to December 23rd, 2022. 
	The color of the trajectory (cyan to magenta) is related to the vessel's speed (slower to faster); grey arrows indicate the direction of the vessel when entering and exiting the Strait of Otranto.
	The orange star in the smaller panel represents the
    vessel's
    long-term predicted position at the acquisition time of the Sentinel-1 SAR image recorded over the strait on December 21st; the SAR image is shown in  Fig.~\ref{fig:adriatic-sea-sar}.}
	\label{fig:adriatic-sea-full-trajectory}
	
\end{figure*}

\begin{figure*}[!t]

	\centering
	\includegraphics[width = .9\textwidth]{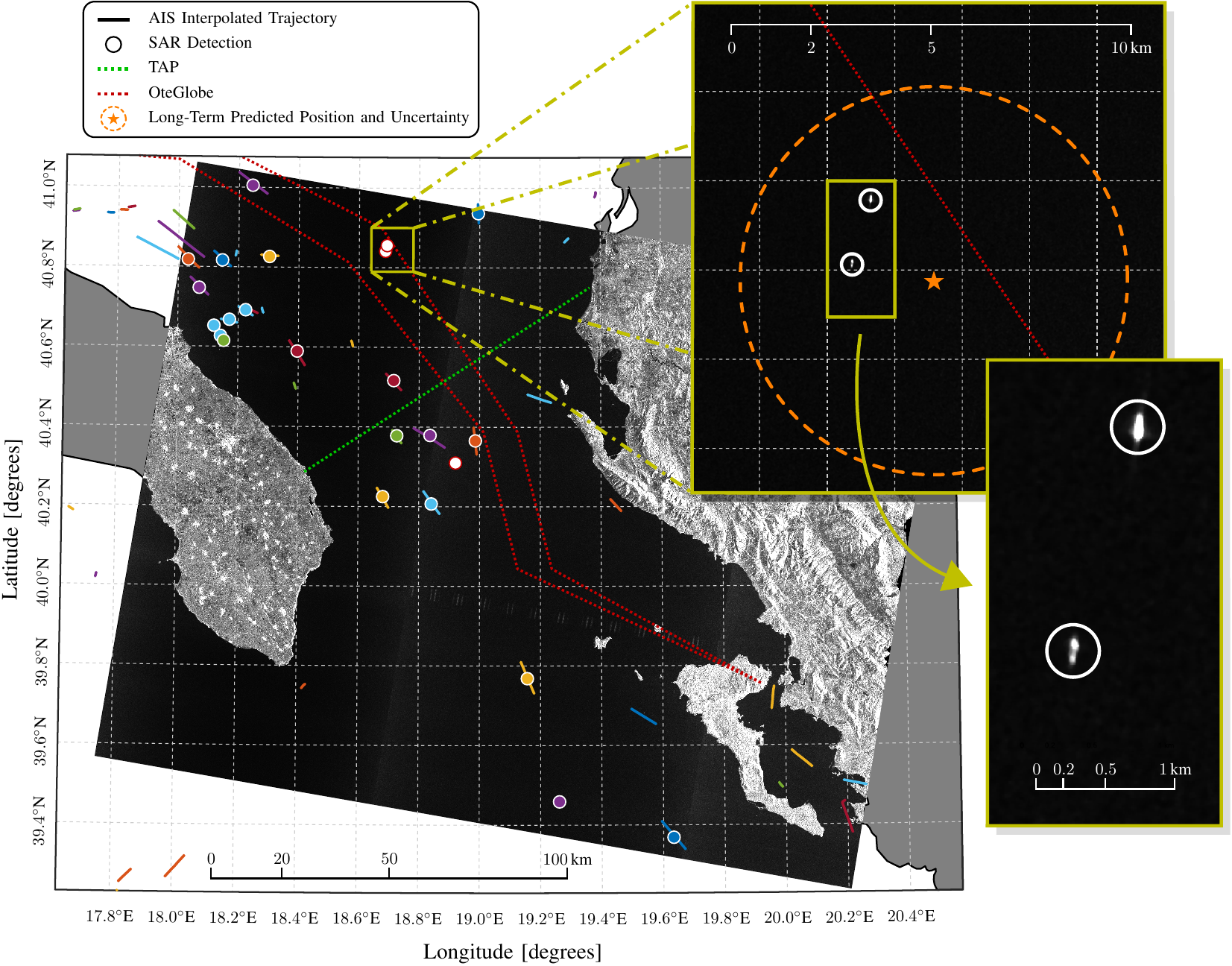}

	\caption{Adriatic Sea use case. Sentinel-1 SAR image acquired over the Strait of Otranto on December 21st, 2022.
	Colored solid lines are vessels' trajectories obtained by interpolating the AIS positions provided by the vessels themselves.
	Circles are ships as detected within the SAR image: the
	color of the circle identifies the detection as coming from a vessel whose AIS trajectory has the same color; a white circle indicates that the detection is not associated with any AIS trajectory.
	Top-right panel reports the
    vessel-of-interest's
    long-term predicted position and its uncertainty with an orange star and dashed circle, respectively.}
	\label{fig:adriatic-sea-sar}
	
\end{figure*}

On October 20th, 2022, the news reported that the south segment of the SHEFA-2 cable, connecting the Shetland Islands to the
UK mainland via the Orkney Islands, was cut; the north segment of the same underwater cable, connecting the Shetland Islands to the Faroe Islands, was also damaged a week earlier, thus causing a major communication outage on the islands for several days~\cite{shetland-accident-press-1}.
The area where the cable lies is interested in an intense fishing activity conducted by multiple trawlers.
It is therefore widely thought that the damage to the SHEFA-2 cable was accidentally caused by a fishing vessel, as also documented in the past, and not the result of a sabotage~\cite{shetland-accident-press-2}.
From an analysis of the AIS data collected in the region several days before the reported accident, it results that multiple boats were engaged in fishing activities.
The trajectory of one of these fishing boats is reported in Fig.~\ref{fig:full-trajectory-shetland}, and demonstrates that the path of the underwater cable was crossed several times in few days.
Over the same period, a cargo vessel was loitering in the area crossing the cable's path few times; however, it is also reported that this vessel left the area few days before the accident.

Although the damage of the SHEFA-2 cable is widely considered to be accidental, it is a clear example of how a malicious coordinated attack targeting multiple links of a network could isolate a territory, thus causing major disservices.

\subsection{Adriatic Sea}
\label{subsec:adriatic}
At the end of 2022, several Italian news agencies reported the anomalous behavior of a
large vessel entering the Strait of Otranto, Adriatic Sea, and stopping over the Trans Adriatic Pipeline (TAP) and the underwater cable OteGlobe for a few hours~\cite{adriatic-sea-press}.
Figure~\ref{fig:adriatic-sea-full-trajectory} shows
its full trajectory
as reported by the AIS from December 19th to December 23rd, and a detail of its route in the Strait of Otranto from the evening of December 20th to the afternoon of December 22nd.
The illustration also reports the paths of the TAP and the OteGlobe (as well as other UCIs in the area) as reported by open sources, which, however, may deviate from their actual routes.

The dashed black lines highlight a gap of more than 21 hours in the AIS data available for the vessel of interest on December 21st.
On this same day, a Sentinel-1 SAR image was acquired over the strait about 3 hours after the latest available AIS position.
Given
this large time interval,
the vessel's position
at the SAR acquisition time is inferred by means of a long-term prediction tool for non-maneuvering targets that uses the Ornstein-Uhlenbeck
process to model the ship's kinematic~\cite{MilBraBryWil:J16,MilBraWil:J16};
the vessel-of-interest's
long-term predicted position is depicted as an orange star, and its uncertainty is represented by an orange circle.

The SAR image is reported in Fig.~\ref{fig:adriatic-sea-sar} along with all the detected ships in the area, represented by circles, and all the vessels' trajectories as reported by the AIS; as mentioned in Section~\ref{subsec:nord-stream}, the color of a circle matches the color of the AIS trajectory which it is associated to, while white circles represent SAR detections that are not associated to any AIS trajectory.
Long-term predicted position and
uncertainty of the vessel of interest are reported in the top-right panel of Fig.~\ref{fig:adriatic-sea-sar}.
Interestingly, the dashed orange circle representing the uncertainty of the prediction encloses two detected vessels that appear to be stationary in close proximity of the underwater cable OteGlobe; these vessels are not associated with any other AIS trajectories, thus suggesting that one of them might actually be 
the large vessel reported by the news agencies.

\section{Conclusion}
\label{sec:conclusion}
Underwater critical infrastructures (UCIs) are fundamental in many fields of everyday life, ranging from 
telecommunications to energy, economy, and finance. Following the explosions affecting the Nord Stream gas pipelines, 
on September 26th, 2022, there has been increasing attention by authorities and international agencies, toward the 
improvement of UCIs resilience. In this article, we have shown how
analyses of the data from automatic 
identification system (AIS) and satellite data, e.g., Synthetic Aperture Radar (SAR), in conjunction with contextual information, such as 
bathymetry and weather data, can
highlight anomalous and suspicious behaviors. Moreover, we have presented how Artificial Intelligence (AI) and 
Information Fusion (IF) can fuse information generated from a large variety of assets, e.g., AIS and SAR, and use the learned knowledge to anticipate future behaviors and/or identify threats and critical situations towards UCIs.

\section*{Acknowledgement}
The authors would like to thank Dr. Alvarez for his hints and the useful discussions about the operational problem, and MarineTraffic for providing the real-world AIS data set used in Figs.~\ref{fig:adriatic-sea-full-trajectory}-\ref{fig:adriatic-sea-sar}. 

\balance
\bibliographystyle{IEEEtran}
\bibliography{IEEEabrv,references}

\end{document}